\documentclass[twocolumn]{revtex4-1}
\usepackage{graphicx}

\newcommand{\chem}[1]{\ensuremath{\mathsf{#1}}}       

\begin{document}
\title{The Promise of Quantum Simulation}
\author{Richard P. Muller}
\email{rmuller@sandia.gov}
\author{Robin Blume-Kohout}
\email{rjblume@sandia.gov}
\affiliation{Center for Computing Research, Sandia National
Laboratories, Albuquerque, NM, 87185-1322}
\begin{abstract}
Quantum simulation promises to be one of the primary application of quantum computers, should one be constructed. This article briefly summarizes the history quantum simulation in light of the recent result of Wang and coworkers demonstrating calculation of the ground and excited states for a HeH+ molecule, and concludes with a discussion of why this and other recent progress in the field suggests that quantum simulation of quantum chemistry has a bright future.
\end{abstract}
\maketitle

In 1982, the physicist Richard Feynman noted that it is very hard for 
a classical computer to simulate some quantum mechanical systems -- but
that if ``the computer itself be built of quantum mechanical elements
which obey quantum mechanical laws" \cite{Feynman:1982}, then the
simulation would become far easier.  With this suggestion, Feynman
inaugurated the field of quantum computing, and introduced the idea that
quantum computers change the boundaries of computational complexity.

Simulation of quantum physical systems is thus the first ``killer app"
of quantum computing.  Since then, several other fast quantum algorithms
have emerged, notably Peter Shor's algorithm for integer
factorization and discrete logarithms \cite{Shor:1994} and Lov Grover's
algorithm for unstructured search \cite{Grover:1996}.  But quantum simulation
remains one of the primary applications of a quantum computer, should
one be constructed.  It is arguably the most intriguing and potentially
valuable of all the known quantum algorithms, with potential impact in
chemistry \cite{Georgescu:2014}, materials science \cite{Georgescu:2014},
and elementary particle physics \cite{Jordan:2012}.

The canonical quantum simulation algorithm uses the same subroutine as
Shor's integer factorization algorithm -- the quantum Fourier transform
-- but instead of using it to reveal the period of a modular function, 
simulation uses it to estimate the energies of a Hamiltonian.  It does
this using Kitaev's algorithm for quantum phase estimation \cite{Kitaev:1995},
which efficiently reveals the phase (and thus the eigenvalue), of an
eigenvector of a unitary operation.  Abrams and Lloyd
\cite{Abrams:1997, Abrams:1999} showed that a broad class of Hamiltonians
could be efficiently simulated (and thus probed using phase estimation) using
a Suzuki-Trotter expansion.  They applied their results to the Hubbard model
and suggested that these techniques would lead to an exponential speed-up
over classical computing resources.  Somma and coworkers \cite{Somma:2002}
extended quantum simulation to many more systems by showing that the 
Jordan-Wigner transformation \cite{Jordan:1928} could be used to map the
creation/annihilation operators that often define quantum mechanical
Hamiltonians into N-qubit operators.  This paved the way for quantum
simulation of quantum chemistry.

Quantum chemistry addresses the problem of describing the electronic
structure of molecules and materials, which provides information about
how they dissociate, react, absorb light, and interact with other
molecules and materials. Typically, quantum chemical methods describe
molecular electronic states as products of single-electron states
generated from a variety of effective Hamiltonians, including
Hartree-Fock \cite{Roothaan:1951} and density functional theory
\cite{Hohenberg:1964, Kohn:1965}.  A huge variety of interesting
chemical systems are well-described either by these product states
or by simple perturbations of them. However, these methods break down
in systems involving strongly correlated electrons.
For these molecules, quantum chemists fall back on brute force methods,
the most accurate (and arduous) of which is full configuration interaction (CI) 
\cite{Lowdin:1955}.
 
Full CI considers the many-body Hamiltonian formed by all the possible
ways of distributing $N$ electrons into $M$ one-electron states.  Because
the number of such configurations grows combinatorially with
$N$ and $M$, full CI computations require [classical] computational resources
that scale exponentially with the molecule's size.  This kind of scaling
rapidly surpasses the capacity of any existing computer, and so quantum
chemists have developed a variety of ``truncated" approximations to full CI.
Typically, these involve configurations that can be generated using just a few
excitations from the ground state electronic configuration.  Unfortunately,
although truncated CI wave functions can be evaluated with only polynomial
resources, they are known to lack properties, such as size consistency 
\cite{Bartlett:1981}, that are required for quantitative chemical predictions.
Related perturbational approaches, such as coupled-cluster approaches
\cite{Cizek:1966}, include size consistency, but often lack strict
variational bounds on the energy.  So, although both truncated CI and
coupled cluster methods often give excellent energies in practice, 
they are incomplete solutions to chemical simulation because they lack
certain properties of the full CI method.  For the chemical problems
where these methods fail, fully quantum simulation a la Feynman might
be the only viable approach.

In 2005, Aspuru-Guzik and coworkers \cite{Aspuru-Guzik:2005} applied
iterative phase estimation and other techniques from quantum information
theory to the full CI problem.  Although their approach scales polynomially for
the full CI problem, the exact scaling of their algorihtm was unclear in 2005.
However, their approach suggested that even modest quantum resources could
potentially enable full CI simulations that outperform the largest, fastest
supercomputers.

The first experimental demonstration of quantum full CI algorithm came
5 years later, when a photonic quantum information processor \cite{Lanyon:2010}
calculated the ground and electronically excited states for a minimal 
basis set description of \chem{H_2} using 20-bit iterative phase 
estimation.  Although this proof-of-principle computation did not
provide any added insight into the nature of the \chem{H_2} bond,
it demonstrated conclusively that the quantum phase estimation algorithm
could produce accurate bonding and excitation energies -- even in the face
of the noise and decoherence inherent to imperfect physical qubits.

In this issue of ACS Nano., Wang and coworkers \cite{Wang:2015} 
demonstrate a quantum calculation of the ground and excited
states for a \chem{HeH^+} molecule, which is isoelectronic with
\chem{H_2}, but has differently-charged nuclei. Wang et al.
report the highest precision achieved to date in the quantum simulation
of molecular energies; they surpass chemical precision by 10 orders of 
magnitude.  Their quantum simulation was performed
using two qubits from a diamond NV center, and represents the first
implementation of the full CI algorithm on a solid-state qubit. As with
the earlier demonstration \cite{Lanyon:2010}, this result is most notable
for verifying that these algorithms can be successfully implemented on
actual qubits, rather than just in theoretical idealizations.

The principle of quantum chemistry simulation has been proved.  The next
urgent question is how it will scale -- how many qubits and how
much time will be required to apply quantum simulation algorithms
to larger, more general molecules.  The time requires scales with $N$, 
the number of basis functions used to describe the atomic orbitals that
comprise the molecule (adequate basis sets require 5-20 basis functions
per atom), and/or $Z$, the maximum nuclear charge in the molecule.  
The initial analysis last year \cite{Wecker:2014} suggested
that quantum simulation algorithms might require $\mathcal{O}(N^9)$
clock cycles.  Although this is a huge advance, in-principle, over the
$\mathcal{O}(e^N)$ scaling of exact classical simulation, it remains
prohibitively intractable in practice.  However, more careful analyses
of the Suzuki-Trotter expansion have produced steady and rapid 
improvements in this scaling, first to $\mathcal{O}(N^7)$
\cite{Hastings:2015}, then to $\mathcal{O}(N^{5.5})$ \cite{Poulin:2014},
and, most recently, to $\mathcal{O}(N^3Z^{2.5})$ \cite{Babbush:2015}. 

The best scaling known today suggests that, given a handful of good qubits,
quantum simulation could produce practical, useful results.  A closer
look, however, suggests a less optimistic picture in the near future.
Molecules with $N=1000$ are routinely analyzed on laptop computers
using density functional theory (DFT), the workhorse method of chemistry and
materials science.  The most optimistic scaling given above suggests that
a quantum computer would need at least $10^9$ operations to
match what DFT can do on a laptop.  Because qubits have far higher error rates
than classical computers, such a large computation would absolutely demand
quantum error correction \cite{Shor:1995}.  Error correction imposes massive
overhead in time, number of qubits, and complexity, 
because small quantum rotations have to be compiled into $H$, $S$, and 
$T$ gates \cite{Kitaev:2002}, some of which must be exhaustively distilled
from noisy resources \cite{Bravyi:2005}.  These considerations suggest that
practical quantum simulation of meaningful molecules is substantially more
challenging than it appears at first -- and perhaps infeasible in the near term.

Perhaps surprisingly, we remain optimistic.  Although the demonstrations by
Wang et al \cite{Wang:2015} and others are a long way from practical utility,
they demonstrate that the principle is sound.  It is widely believed that
further algorithmic improvements are possible -- after all, just in the past
two years we have seen the time scaling drop from $\mathcal{O}(N^9)$ to 
$\mathcal{O}(N^3Z^{2.5})$. Moderate improvements in scaling could
dramatically change feasibility, and algorithms that avoid the
Suzuki-Trotter expansion \cite{Cleve:2009,Berry:2015,Babbush:2015a,Babbush:2015b}
appear promising.  
Finally, we observe that some of the
most promising candidates for near-term ``quantum supremacy" (practical speedups
over existing classical computers) are robust quantum simulation algorithms
that may not require error correction, either by using device noise to mimic
real-world noise in the simulated system, or by leveraging shallow quantum
circuits that run so quickly that errors do not accumulate.  
Analog quantum simulation \cite{Georgescu:2014} is one such candidate,
and has been used to mimic Hubbard \cite{Greiner:2002} and spin \cite{Friedenauer:2008}
Hamiltonians.  Analog methods are not currently known for chemical or molecular systems,
but these systems \emph{can} be addressed within the variational eigensolver approach \cite{Peruzzo:2014}, which leverages a small quantum computer to evaluate the energy
terms of a parametric wave function that is varied by an associated classical
computer-driven optimization.  We believe that between the near-term promise of these 
non-traditional algorithms, and steady progress toward the long-term goal of digital
quantum simulation, quantum simulation of quantum chemistry has a bright future.

Sandia is a multiprogram laboratory operated by Sandia Corporation, a
Lockheed Martin Company, for the United States Department of Energy's
National Nuclear Security Administration under Contract
DE-AC04-94AL85000. This document is the unedited Author'€™s version of a
Submitted Work that was subsequently accepted for publication in ACS
Nano, copyright 2015 American Chemical Society after peer review. To
access the final edited and published work see \cite{Muller:2015}.

\bibliographystyle{unsrt}
\bibliography{ACSNanoPerspectives2}

\end{document}